\documentstyle[12pt,aaspp4,flushrt,psfig]{article}
\def\etal{{et al.}}
\begin{document}

\title{L Dwarfs Found in Sloan Digital
Sky Survey Commissioning Imaging
Data\footnote{Based on observations obtained with the
Sloan Digital Sky Survey, and with the Apache Point Observatory
3.5-meter telescope, which is owned and operated by the Astrophysical
Research Consortium}}

\author{Xiaohui Fan\altaffilmark{\ref{Princeton}},
G. R. Knapp\altaffilmark{\ref{Princeton}},
Michael A. Strauss\altaffilmark{\ref{Princeton}},
James E. Gunn\altaffilmark{\ref{Princeton}}, 
Robert H. Lupton\altaffilmark{\ref{Princeton}}, 
\v{Z}eljko Ivezi\'{c}\altaffilmark{\ref{Princeton}},
Constance M. Rockosi\altaffilmark{\ref{Chicago}}, 
Brian Yanny\altaffilmark{\ref{Fermilab}}, 
Stephen Kent\altaffilmark{\ref{Fermilab}},
Donald P. Schneider\altaffilmark{\ref{PennState}}, 
J. Davy Kirkpatrick\altaffilmark{\ref{IPAC}},
James Annis\altaffilmark{\ref{Fermilab}},
J. Brinkmann\altaffilmark{\ref{APO}}, 
Istv\'an Csabai\altaffilmark{\ref{JHU},\ref{Eotvos}},
Masataka Fukugita\altaffilmark{\ref{CosmicRay},\ref{IAS}},
G. S. Hennessy\altaffilmark{\ref{USNO}},
Robert B. Hindsley\altaffilmark{\ref{USNO}},
Takashi Ichikawa\altaffilmark{\ref{Tohoku}},
D.Q. Lamb\altaffilmark{\ref{Chicago}},
Timothy A. McKay\altaffilmark{\ref{Michigan}},
Jeffrey A. Munn\altaffilmark{\ref{Flagstaff}},
Sadanori Okamura\altaffilmark{\ref{UTokyo}},
A. George Pauls\altaffilmark{\ref{Princeton}},
Jeffrey R. Pier\altaffilmark{\ref{Flagstaff}},
Ron Rechenmacher\altaffilmark{\ref{Fermilab}},
Alexander S. Szalay\altaffilmark{\ref{JHU}},
Douglas L. Tucker\altaffilmark{\ref{Fermilab}},
Donald G. York\altaffilmark{\ref{Chicago}} (the SDSS Collaboration)
}

\newcounter{address}
\setcounter{address}{2}
\altaffiltext{\theaddress}{Princeton University Observatory, Princeton, NJ 08544
\label{Princeton}}
\addtocounter{address}{1}
\altaffiltext{\theaddress}{University of Chicago, Astronomy \& Astrophysics
Center, 5640 S. Ellis Ave., Chicago, IL 60637
\label{Chicago}}
\addtocounter{address}{1}
\altaffiltext{\theaddress}{Fermi National Accelerator Laboratory, P.O. Box 500,
Batavia, IL 60510
\label{Fermilab}}
\addtocounter{address}{1}
\altaffiltext{\theaddress}{Department of Astronomy and Astrophysics,
The Pennsylvania State University,
University Park, PA 16802
\label{PennState}}
\addtocounter{address}{1}
\altaffiltext{\theaddress}{Infrared Processing and Analysis Center,
100-22 California Institute of Technology,
Pasadena, CA 91125
\label{IPAC}}
\addtocounter{address}{1}
\altaffiltext{\theaddress}{Apache Point Observatory, P.O. Box 59,
Sunspot, NM 88349-0059
\label{APO}}
\addtocounter{address}{1}
\altaffiltext{\theaddress}{
Department of Physics and Astronomy, The Johns Hopkins University,
   3701 San Martin Drive, Baltimore, MD 21218, USA
\label{JHU}
}
\addtocounter{address}{1}
\altaffiltext{\theaddress}{Department of Physics of Complex Systems,
E\"otv\"os University,
   P\'azm\'any P\'eter s\'et\'any 1/A, Budapest, H-1117, Hungary
\label{Eotvos}
}
\addtocounter{address}{1}
\altaffiltext{\theaddress}{Institute for Cosmic Ray Research, University of
Tokyo, Midori, Tanashi, Tokyo 188-8502, Japan
\label{CosmicRay}}
\addtocounter{address}{1}
\altaffiltext{\theaddress}{Institute for Advanced Study, Olden Lane,
Princeton, NJ 08540
\label{IAS}}
\addtocounter{address}{1}
\altaffiltext{\theaddress}{U.S. Naval Observatory, 
3450 Massachusetts Ave., NW, 
Washington, DC  20392-5420
\label{USNO}}
\addtocounter{address}{1}
\altaffiltext{\theaddress}{Astronomical Institute,
Tohoku University,
Aoba, Sendai 980-8578
Japan
\label{Tohoku}}
\addtocounter{address}{1}
\altaffiltext{\theaddress}{University of Michigan, Department of Physics,
	500 East University, Ann Arbor, MI 48109
\label{Michigan}}
\addtocounter{address}{1}
\altaffiltext{\theaddress}{U.S. Naval Observatory, Flagstaff Station, 
P.O. Box 1149, 
Flagstaff, AZ  86002-1149
\label{Flagstaff}}
\addtocounter{address}{1}
\altaffiltext{\theaddress}{Department of Astronomy and Research Center for the Early Universe,
	School of Science, University of Tokyo, Hongo, Bunkyo,
Tokyo, 113-0033 Japan
\label{UTokyo}}
\newpage
\begin{abstract}
This paper describes the discovery of seven dwarf 
objects of spectral type `L' (objects
cooler than the latest M dwarfs) in commissioning imaging data taken by the 
Sloan Digital Sky Survey (SDSS).  Low-resolution spectroscopy shows that
these objects have spectral types from L0 to L8.
Comparison of the SDSS and 2MASS photometry
for several of these objects indicates
the presence of significant opacity at optical wavelengths, perhaps due to
atmospheric dust.  This comparison also demonstrates the high astrometric
accuracy (better than $1''$ for these faint sources) of both surveys.

The L dwarfs are shown to occupy a distinctive region of color-color
space as measured in the SDSS filters, which should enable their identification
in a straightforward way. This should lead eventually to a complete sample of many
hundreds of these low mass objects, 
or about one per 15 square degrees to $i' \approx 20$, in the 
complete SDSS data set.

\end{abstract}

\section{Introduction}

The long search for substellar objects, or brown dwarfs, has finally been 
successful.  The last three years have seen spectacular advances in this
field, from the discovery of low-luminosity objects
which are companions to
nearby stars (Nakajima \etal\ 1995; Goldman
\etal\ 1999),
are in young clusters (Zapatero-Osorio 1998), or are in
the field (Ruiz, Leggett \&
Alard 1997; Delfosse \etal\ 1997;
Kirkpatrick \etal\ 1999, hereafter K99). Many of these dwarfs,
of spectral type `L', have detectable lithium and are therefore
brown dwarfs.  At the
same time, new planetary systems around nearby stars (Marcy \& Butler 1998,
Queloz \etal\ 1998, Mayor \etal\ 1998, Butler
\& Marcy 1998) have been found, containing planets 
whose masses are several times that of Jupiter. 
The companion object Gl 229B (Nakajima \etal\ 1995) is cool enough
($\rm T_{eff} \sim 900$K) that its atmosphere resembles that of
a giant gas planet, with strong absorption bands of methane.  Very
recently, several isolated objects with properties similar to those of
Gl 229B, the so-called `methane', or `T', dwarfs 
(objects with methane bands at K), have been found by deep optical
(Strauss \etal\ 1999; Pier \etal\ 1999; Tsvetanov \etal\ 1999) and
infrared (Burgasser \etal\ 1999; Cuby \etal\ 1999) surveys.  Because
brown dwarfs never reach an equilibrium, or ``main sequence'', state, 
there is no one-to-one correspondence between the photometric or spectroscopic
properties and the stellar mass.  The
large number and range of new discoveries in the field are well summarized
in the recent volumes edited by Rebolo, Mart\'{\i}n
\& Zapatero-Osorio (1998) and Marley (1999), and the reviews by
Oppenheimer, Kulkarni \& Stauffer (1999) and Liebert (1999).

Substellar objects have until recently
been extremely difficult to find: they are very
low luminosity ($\rm < 10^{-3} ~ L_{\odot}$) and very cool ($\rm
T_{eff} \leq 2000$ K), with most (${}>90\%$) of their luminosity emitted at
wavelengths $\rm > 1 ~ \mu m$. 
This means that while
such objects may well be numerically the most common, 
they are among the rarest in magnitude-limited surveys, and large areas
of sky must be covered to faint magnitude limits to find them.
Wide area surveys at optical wavelengths, made with 
Schmidt photographic plates, have found a few of these objects in the
field and in young clusters (Rebolo, Mart\'{\i}n, \& Zapatero-Osorio 1998
and references therein). 
However, the discovery of most new
field objects has come with the new generation of sensitive, all-sky,
near-infrared surveys, DENIS and 2MASS (Delfosse \etal\ 97; K99).


K99 discuss the selection of candidate cool,
low luminosity objects using 2MASS J, H and $\rm K_s$ photometry,
together with optical upper limits from photographic surveys.  They
have obtained spectra for a large number of these candidates, 
and have established the 
continuation of the spectral classification sequence to types later than
the coolest M dwarfs. 
These `L' dwarfs (cf. Mart\'\i{}n \etal\ 1997) form a spectral class
distinguishable from M dwarfs by the eventual 
disappearance of the strong TiO and 
VO bands (implying condensation into solid particles in the atmosphere)
and the appearance of bands of metal hydrides and lines of alkali metals.  

The present paper discusses the discovery of seven new field `L' dwarfs
from the commissioning imaging data of the Sloan Digital Sky Survey
(SDSS).  We have no conclusive data to say whether or not
these objects are brown dwarfs, which would require a direct measure
of their mass, and sidestep this issue by confining our discussion to
their spectral and photometric properties. 


The next section describes the SDSS imaging data in which the candidate L
dwarfs were found.  The selection of the L dwarf candidates is discussed in
Section 3, and the spectroscopy in Section 4.  Section 5 describes the 
broad-band characteristics of these objects at optical and near-infrared 
wavelengths, using SDSS and 2MASS photometry.  Section 6 discusses the sky
distribution of objects with L dwarf colors and a very preliminary estimate
of their areal density.
The conclusions are given in Section 7.

\section{Photometric Observations}

The photometric observations were made in Fall 1998 to Spring
1999 with the SDSS 
2.5 meter telescope and imaging camera at the Astrophysical Research
Consortium's Apache Point Observatory (APO), New Mexico.  The SDSS is
described by Gunn \& Weinberg
(1995) and York \etal\ (2000\footnote{see also {\tt
http://www.astro.princeton.edu/PBOOK/}}).  The telescope
is a modified f/5 Ritchey-Chr\'etien optical system with a large
secondary mirror and first and second corrector lenses 
above and below the primary
mirror, which produces a $3^\circ$ distortion-free field (Siegmund
\etal\ 2000\footnote{see also {\tt
http://www.astro.princeton.edu/PBOOK/telescop/telescop.htm}}).  

The imaging camera (Gunn \etal\ 1998) contains two arrays of CCD detectors.  The imaging 
array is a mosaic of 30 2048$\times$2048 SITe CCDs with 24$\rm \mu m$
($\rm 0.4''$ on the sky) pixels.  The CCDs are arranged in six
columns of 5 CCDs.  Each column 
observes the sky through five broad-band filters [$u'$, $g'$, $r'$,
$i'$ and $z'$], with effective wavelengths [3540\AA{},
4760\AA{}, 6280\AA{}, 7690\AA{}, and 9250\AA{}], covering the entire atmospheric window from the 
atmospheric ultraviolet cutoff to the silicon red sensitivity cutoff
(Fukugita \etal\ 1996).
The total integration
time per filter is 54.1 seconds, and the expected survey
depths in the five filters (5$\sigma$ detection of a point source with $\rm 1''$
FWHM images) are [$\rm 22.3$, $\rm 23.3$, $\rm 23.1$, $\rm 22.5$
and $\rm 20.8$] (Gunn \etal\ 1998).
The camera also 
contains two additional arrays of neutral-density-filtered, $r'$ CCDs
of 2048$\times$400 pixels, which saturate at $r'\sim 8$ and cover
the dynamic range between the photometric CCDs and existing astrometric catalogues, 
providing the astrometric calibration for objects detected by the
photometric CCDs (Gunn \etal\ 1998; Pier \etal\ 2000\footnote{see also {\tt
http://www.astro.princeton.edu/PBOOK/astrom/astrom.htm}}).

The photometric data are taken in open-loop time-delay-and-integrate (TDI)
mode at sidereal rate, with a given point in the sky passing through
each of the filters in succession.  This produces a set of six long
continuous imaging scans of the sky, each about $\rm 13'$ wide, and separated
by about 90\% of the CCD width.  
Such a set of imaging data is called a 
{\it strip}; a (filled) {\it stripe} is the combination of two strips
offset by the CCD separation.
The data are read from the imaging camera to disk and tape 
(Petravick \etal\ 1994, 2000\footnote{see also {\tt
http://www.astro.princeton.edu/PBOOK/datasys/datasys.htm}}),
and the tapes sent to Fermilab
for data processing.

The commissioning photometric data on which this paper is based are calibrated by
observations of secondary standard stars in the survey area which were made
with the U.S. Naval Observatory's $\rm 40''$ telescope and a (now
decommissioned) $\rm 24''$ telescope at APO (see Smith et al. 1998;
Tucker et al. 1999).  
As the standard star system was still being developed while these data
were taken, and because of unmodeled small-scale PSF variations in
the mosaic camera data, the 
absolute photometric accuracy of the data discussed in the present paper
is about 5\% - 10\%.  Because of this, we will denote the preliminary SDSS magnitudes we
have measured as $u^*, g^*, r^*, i^*$ and $z^*$,
rather than the notation $u'$, $g'$, $r'$, $i'$, and $z'$ that will be
used for the final SDSS photometric system (and is used in this paper
to refer to the SDSS filters themselves). 

The data were taken 
by parking the telescope at the Celestial Equator and allowing the
sky to drift by.  The scans thus cover 
$\pm 1.25^\circ$ centered on the Celestial
Equator.
Two 
strips were observed in most cases (to make a filled stripe), with northern
(N) and southern (S) offsets.  The particulars of each data run used in this
paper are given in Table 1, which lists the bookkeeping number of each data 
run, whether it is a north or south strip, the right ascension range (J2000),
the date and Modified Julian Day (MJD) of the observation, the approximate area 
covered, and the FWHM of the point spread function (PSF).  
For most of the runs, the PSF is $\rm \sim 1.3''$.  Taking overlaps
into account, these scans cover roughly 600 square degrees. 

The imaging data are processed through a series of automated pipelines
which carry out astrometric and photometric calibrations, flat-field,
bias-subtract, and correct the images for defects, and identify
and measure the properties of all detected objects (Kent
\etal\ 2000; Pier \etal\ 2000;
Tucker \etal\ 2000; Lupton \etal\ 2000\footnote{see also {\tt
http://www.astro.princeton.edu/PBOOK/datasys/datasys.htm}}). 
The SDSS photometric system, and its calibration, are described by 
Fukugita \etal\ (1996) (cf. Equations (1) and (2) in that paper).  
It is based on the $\rm AB_{\nu}$ system of monochromatic
magnitudes,
referenced to an absolute flux scale
in Jy. 

The SDSS uses a modified magnitude scale $\mu(f)$ to handle
the low signal to noise ratio r\'egime, including slightly negative
fluxes $f$ (Lupton \etal\ 1999): 
\begin{equation}
\mu(f) = \mu(0) - a {\rm sinh}^{-1}\left({{f}\over{2 b'}}\right)
\end{equation}
where $a= 2.5 \log e$ and $\rm \mu(0)$ is the normalizing magnitude which sets
the zero point of the magnitude scale.  
The quantity $b'$ is set so that $\mu$ undergoes a transition from
the usual logarithmic to a linear flux scale at roughly 3 times the
sky noise in a PSF aperture. 
The values of $b'$ and $\mu(0)$ vary from run to run, and
the values for runs 94 and 125 are given by Fan
\etal\ (1999a). 
Negative fluxes are represented with values of $\mu > \mu(0)$; 
a detection
at the $5\,\sigma$ level has $\mu \sim 22.4, 23.1, 22.7, 22.1$, and
20.7 
at $u^*, ~ g^*, ~ r^*, i^*$ and $z^*$.

\section{The Selection of L dwarf Candidates in the Photometric Data}

The L dwarfs discussed in this paper were selected from the reduced 
SDSS photometric
data summarized in Table 1 for spectroscopic follow up (described below)
as part of a search for very high redshift quasars, whose results are 
described by Fan \etal\ (1999a,b).  
High redshift quasars ($z\geq 4$) have $g' - r'$ colors several 
magnitudes redder than any ordinary star but also have blue $r'-i'$ colors,
as the Ly$\alpha$ emission
moves into the $r'$ filter (Fan 1999).  Above $z = 4.5$, 
the $i'-z'$ color of quasars is bluer than stars of the same $r'-i'$,
as Ly$\alpha$ moves into the $i'$ filter.  However, several point source objects
were noted in the SDSS data 
on the {\em other} side of the stellar locus,
namely with much {\em
redder} colors in $i^*-z^*$ for their $r^*-i^*$ than ordinary stars.

Figure 1 shows a color-color plot of point-source objects detected in
$r^*, i^*$ and $z^*$ from Run 94 (see Table 1) with  $i^* <20.2$.
Contours of density in color-color space are shown for all except the
outer parts of the stellar locus. Overlaid on this plot
are the colors and spectral
classifications of the eight objects for which spectroscopy
was obtained by the APO 3.5 m telescope (see next section),
as are the colors for an additional SDSS L dwarf observed with the 
Hobby-Eberly Telescope (HET) by Schneider \etal\ (1999). 
Figure 2 shows a color-magnitude diagram
($i^*-z^*$ versus $z^*$) from the same data.  Note that the SDSS
T dwarfs (Strauss \etal\ 1999; Pier \etal\ 1999; Tsvetanov
\etal\ 1999) are redder yet in $i^* - z^*$ than are the L dwarfs. 
While the M dwarf colors lie on the extrapolation of
the stellar locus, 
the L dwarf colors are up to 1 magnitude {\em bluer} in $r^*-i^*$
for a given $i^* - z^*$ than are the 
late M stars, and $\rm {}> 0.5^m$ redder in $i^*-z^*$.  
This region is thus well displaced from the stellar locus and its extrapolation
to the red.  Figure 1 suggests that L dwarfs tend to lie in the
region: 

\begin{eqnarray} 
i^*-z^* &> &1.6 \\
r^*-i^* &>& 1.8
\label{eq:ldwarfcut}
\end{eqnarray}
with the demarcation between `L' and `T' in these colors not yet known. 

Table 2 lists the positions and photometry for the eight objects from Figures 
1 and 2 for which spectra were obtained.  The objects are named by the J2000
coordinates, which are accurate to about $0.1''$ (see Section
5 below).  Next is the
number of the run in which the object was found. 
The asinh magnitudes and their errors are given in Columns 3-7.  
The $i'$ and $z'$ finding charts for these objects 
are shown in Figure 3.

\section{Spectroscopic Observations}

Spectra of the eight objects in Table 2 were obtained using the Double
Imaging Spectrograph (DIS) built by J. Gunn, M. Carr and R. Lupton,
on the ARC 3.5m telescope at the Apache Point Observatory, in
November and December 1998 and March and May 1999.  The DIS has
a transition wavelength of 5350\AA{} between its red and blue sides.
The observations were taken with the low-resolution grating, which gives a
dispersion of 6.2\AA{}/pixel on the blue side and 7.1\AA{}/pixel on the red 
side.  The spectrograph resolution is about 2 pixels, and the joint spectrum 
extends from 4,000\AA{} to 10,000\AA{}.  Exposure times ranged from
1200 to  3600 sec.  Observations
of the F subdwarf standard stars BD+$17^\circ\,$4708 and HD 19445
(Oke \& Gunn 1983) allowed correction for the atmospheric absorption bands
and flux calibration of the spectra.  As some of these observations
were made under non-photometric conditions, the flux calibration was
adjusted to the $i^*$ magnitude of each object.
The wavelength scale was calibrated to about 0.5\AA{} with a 
cubic polynomial fit to the lines of an Ar-He-Ne arc lamp.  The data 
were reduced using the IRAF package (Tody 1993).  Table 3 summarizes
the spectroscopic observations, giving the object name, the date of 
observation, the approximate flux density at 9500\AA{},
and the assigned spectral type, 
found by comparing the spectra with those displayed by K99.

The spectra are shown in spectral sequence in Figure 4, normalizing
the flux densities to a common value at 9500\AA{}.  Figure 5 shows the
spectrum of the brightest L dwarf in this sample, 
SDSS 0539$\,-$0059, with the major features 
marked following the identifications of K99.
The disappearance of the strong VO and, especially, TiO bands below 
about 8000\AA{} is apparent in the spectral sequence in Figure 4,
causing 
the L dwarfs to be bluer in $r^*-i^*$ than the M dwarfs (see Figure 1).
The color temperature of the L dwarfs is lower and their overall 
spectral shape much redder than M dwarfs: hence
the L dwarfs become
increasingly redder in $i^*-z^*$ towards later spectral types.
The main absorption features in the far red spectra
of L dwarfs are due to alkali metals (K, Rb, Na and Cs), metal hydrides
(CaH, CrH and FeH), and strong $\rm H_2O$ absorption in the 
9200 to 9400\AA{} range (K99).  
As discussed by K99, the KI doublet at 7665, 7700 \AA{}
increases in width as one continues down the sequence, becoming many
hundreds of \AA{}ngstroms broad for late L dwarfs. 

\section{Comparison with 2MASS}

The Two-Micron All Sky Survey (2MASS, Skrutskie \etal\ 1997) has scanned the
regions of the sky containing six of the objects in Table 2, including the 
M dwarf.  The  corresponding 2MASS objects and their J, H and $\rm K_s$ 
magnitudes are listed in Table 4. 

Table 5 provides a comparison of the flux densities, listing a set of 
joint optical and infrared colors ordered by spectral type. 
Note that the 2MASS magnitudes are referred to Vega, which is assumed
to be $0^m$ at all wavelengths.
All colors show a general increase towards later spectral
types, with the exception of $r^*-i^*$,  which is bluer for L
dwarfs than it is for M dwarfs. This color also shows very little
correspondence with spectral type at all, although the errors on 
the colors of some stars are large, and may mask a subtle trend. 
The distinctive colors in Table 4 allow L dwarfs to be easily distinguished in
both the SDSS and 2MASS surveys individually; comparison of both surveys
will most directly allow the characterization of the transition
between M and L dwarfs.

The color information between the 2MASS and SDSS surveys is
not redundant.  Figure 6 shows the
broad-band spectrum of L0330$\,-$0025 (L2) compared with black body
curves; the other objects show a similar trend (see Table 5).  
This figure is illustrative only, since cool stars are far from
black bodies; it demonstrates that the L  
dwarfs have a {\em lower} color temperature in the $r^*$ to
$z^*$ regime than in the JHK$_s$ regime, i.e. the 
optical flux densities are depressed relative to the infrared flux
densities, demonstrating the presence of additional opacity at
optical wavelengths.  The likely source is dust, in the form of
condensed TiO and other refractories (Tsuji, Ohnaka \& Aoki, 1996,
1999; Griffith, Yelle \& 
Marley 1998).  Since both 2MASS and SDSS provide precision photometry,
the dust content of the envelopes of M and L dwarfs can be
investigated in detail.  One 
potential source of worry is that because these objects are so red,
the fitting of models to the $z^*$ photometry is critically dependent
on the poorly understood red tail of the $z^*$ response.  Work is
ongoing (Doi \etal\ 2000) to characterize the SDSS system response in detail. 

Table 4 also lists the total offset 
between the 2MASS and SDSS positions.  The positional agreement is
excellent, which is a tribute to the 
astrometry of both surveys (carried out by the United States
Naval Observatory in both cases), given the completely independent
hardware, software and calibration procedures of the two surveys and the fact
that the objects in Table 4 are among the faintest detected in either 
survey.   
This has several interesting implications for future comparison of
2MASS and SDSS.  First, the cross-identification of both surveys should
be very straightforward, and will provide a discriminant against rapidly
moving objects such as asteroids in the 2MASS data 
(this is in general not a severe problem with the SDSS data, see Section 6).
Second, if a source in one survey is not 
matched in the other (and is not an asteroid), 
one can with some confidence set upper limits
(or even 1 or 2$\sigma$ measurements) on the flux densities in the 
non-detected bands by examining the imaging data at the appropriate 
position (as already done by both surveys individually).  
Third, one can 
search for objects which have moved
between the two survey epochs - a proper motion survey over a quarter
of the sky to the survey limiting magnitudes and to a level of a few tens of 
a milliarcsecond per year may be within reach.  

\section{The Sky Density of L Dwarfs in the SDSS}

The L dwarfs described in this paper, together with the L0 dwarf whose
spectrum was measured by the HET (Schneider \etal\ 1999) were selected from
data taken over about 600 square degrees, and thus give a minimum 
rate of detection of L dwarfs by SDSS of about 1 per 75 square degrees
to $i^* = 20.2$.
However, only a small fraction of all the L dwarf candidates in this data
set have been observed spectroscopically, and an upper bound on the
expected SDSS detection rate can be found simply by counting objects
in the ``L dwarf'' region of color-color space (Equations 2 and 3). 
This will work
provided there is no significant contamination of this region of
color-color space by other objects.  We are encouraged by the fact
that we have not yet taken a spectrum of any object in this region of
color-color space that was not an L or T dwarf.
The possible sources of such
contamination are: 
\begin{itemize} 
\item Asteroids, whose proper motion between exposures in different
filters could yield incorrect colors.  However, this motion is looked
for in the analysis of every SDSS object; this removes an important
source of contamination (see Reid et al. 1999)
outside of the stellar locus.
\item Quasars. As the models by Fan (1999) and
the observation of high-redshift quasars by Fan \etal\ (1999a,b) show,
the L dwarf and quasar regions of color-color space do not overlap (at
least for quasars with $z < 5.5$), and all objects in this region
whose spectra have been measured are indeed L dwarfs.
\item M dwarfs.  We have seen that 
the L dwarfs separate well in $r^*-i^*$, $i^*-z^*$ color-color
space from the observed M7 dwarf, but the spectral type where this
color transition occurs is not yet determined.
\item Heavily reddened or peculiar stars.  SDSS commissioning
imaging scans over the
dusty, star-forming regions in Orion in fact yield a clustered
population of objects in the L dwarf region of the color-color
diagram.  We are currently investigating their properties, and will
discuss them in detail in a future publication. Otherwise, the SDSS
photometry, being generally taken at high latitudes, appears to separate
(possibly) very late M, L and T dwarfs from all other point-source objects.
\end{itemize}

Thus the count of reliably detected point sources in
the L dwarf region of color space gives at 
least an upper bound, and probably a reasonable
estimate, 
of how many L dwarfs the SDSS can be expected to find.

Several of the data runs listed in Table 1 were examined for faint objects
with L dwarf colors; the results are listed in Table 6. 
These very preliminary estimates suggest that the SDSS will detect
about 1 L dwarf per 15 square degrees to $i^* = 20.2$, comparable to
the 2MASS detection rate (K99), and suggesting that the SDSS will
detect 
perhaps up to 1000 of these objects over the 10,000 square degrees of
the survey.  The SDSS and 2MASS data sets are likely to
be both overlapping and complementary: 2MASS is more sensitive to
the very coolest dwarfs, while SDSS will be more sensitive to the hotter
objects.  Together, these surveys will allow the population below about M5 
to be characterized.  

\section{Conclusions}

This paper discusses members of the new spectral class of `L' dwarfs 
(K99) found in commissioning imaging data of the Sloan Digital Sky Survey and 
confirmed by spectroscopy from 5000\AA{} to 1$\rm \mu m$ with the
Double Imaging Spectrograph on the 
ARC 3.5 m telescope at Apache Point, New Mexico. 

\begin{enumerate}

\item
Seven `L' dwarfs were spectroscopically confirmed, and an eighth has
been confirmed at the HET (Schneider \etal\ 1999).  The expected detection
rate for L dwarfs has also been estimated by counting point source objects 
in the SDSS imaging data with `L' dwarf colors.  These very preliminary
estimates show that the detection rate for L dwarfs in the SDSS will
likely lie somewhere between a lower limit of 1 per 75 square degrees and 
an upper limit of 1 per
15 square degrees to $i^* = 20.2$.  
If the higher number turns out to be correct, the 
SDSS detection rate is similar to that of recent, sensitive near-infrared
surveys such as 2MASS.

\item
L dwarfs are 
redder in $i^*-z^*$ than the latest M dwarfs and become redder 
towards later spectral types, as expected from the decrease in color
temperatures shown by the spectra.  However, they are significantly
{\em bluer} than the late M dwarfs in $r^* - i^*$ because of the diminished
VO and TiO absorption.  This makes them easy to identify
in the SDSS $r^*-i^*$,$i^*-z^*$ color-color diagram, and SDSS data are
likely to lead to a detailed
characterization of the number density of the latest M dwarfs and the
M dwarf $\rightarrow$ L dwarf transition.

\item
The SDSS $r^*, ~i^*$ and $z^*$ flux densities show that the L dwarfs
have lower color temperature at optical than at near infrared wavelengths,
consistent with the presence of extra optical opacity in the
envelopes, perhaps due to dust.  
Thus the
combination of the SDSS and 2MASS photometry may allow the dust to be 
characterized in L dwarfs and in the yet cooler T dwarfs.

\item
The positions of these faint objects as measured by 2MASS and SDSS agree
to better than $1''$, a tribute to the astrometric integrity of both
surveys.  Positional correlation of the data sets and searches for
proper motions will be straightforward.

\end{enumerate}

\acknowledgements
The Sloan Digital Sky Survey (SDSS) is a joint project of The
University of Chicago, Fermilab, the Institute for Advanced Study, the
Japan Participation Group, The Johns Hopkins University, the
Max-Planck-Institute for Astronomy, Princeton University, the United
States Naval Observatory, and the University of Washington.  Apache
Point Observatory, site of the SDSS, is operated by the Astrophysical
Research Consortium.  Funding for the project has been provided by the
Alfred P. Sloan Foundation, the SDSS member institutions, the National
Aeronautics and Space Administration, the National Science Foundation,
the U.S. Department of Energy, and the Ministry of Education of Japan.
The SDSS Web site is at {\tt http://www.sdss.org/}.
We thank Karen Gloria and Russet McMillan for their usual expert
help with the APO 3.5m observations.
This publication makes use of data from the Two Micron All Sky Survey,
which is a joint project of the University of Massachusetts and the Infrared 
Processing and Analysis Center, funded by the National Aeronautics and 
Space Administration and the National Science Foundation.
XF and MAS acknowledge additional
support from Research Corporation, NSF grant AST96-16901, the
Princeton University Research Board, and an Advisory Council
Scholarship. GRK is grateful for support from Princeton University and
from the NSF via grant AST96-18503.  DPS thanks the NSF for support
from grants AST95-09919 and AST99-00703.

\begin{table}
\scriptsize
\centerline{Table 1.  SDSS Photometric Commissioning Observations on the Celestial
Equator}
\bigskip
$$\vbox{
\tabskip 1em plus 2em
\halign to \hsize{\hfil$\rm #$\hfil& &   \hfil$\rm#$\hfil \cr
Run& N/S& RA~range& RA~range& Date& MJD& Area& PSF\cr
Number& & (J2000)& (J2000)& & 240000.0+& (sq. deg.)& (arcsec)\cr
& & & & & & & \cr
77& N& 12^h36^m31^s \rightarrow 16^h39^m03^s& 
189.1 \rightarrow 249.8& 98/06/27& 50991& 82& 1.3\cr
& & & & & & & \cr
85& S& 13^h29^m50^s\rightarrow 18^h42^m26^s& 
202.5\rightarrow280.6& 98/06/27& 50992& 106& 1.3\cr
& & & & & & & \cr
94& N& 22^h16^m00^s\rightarrow03^h49^m33^s& 
334.0\rightarrow 57.4& 98/09/19& 51075& 110& 1.5\cr
& & & & & & & \cr
109& S& 02^h20^m43^s\rightarrow04^h35^m28^s& 
35.2\rightarrow68.9& 98/09/22& 51078& 45& 1.3\cr
& & & & & & & \cr
125& S& 23^h18^m02^s\rightarrow05^h17^m21^s& 
349.5\rightarrow79.3& 98/09/25& 51081& 122& 1.6\cr
& & & & & & & \cr
211& S& 02^h46^m31^s\rightarrow06^h43^m40^s& 
41.6\rightarrow100.9& 98/10/29& 51115& 80& 
1.3-1.4\cr
& & & & & & & \cr
259& N& 00^h29^m05^s\rightarrow06^h27^m47^s& 
7.3\rightarrow89.7& 98/11/17& 51134& 121&
2.0-1.0\cr
& & & & & & & \cr
273& S& 00^h42^m12^s\rightarrow06^h10^m22^s& 
10.6\rightarrow 92.6& 98/11/19& 51136& 111& 
1.6-1.4\cr
& & & & & & & \cr
745& N& 10^h37^m36^s\rightarrow 16^h50^m24^s& 159.4\rightarrow 252.6&
99/03/20& 51257& 126& 1.6-1.2\cr
& & & & & & & \cr
752& S& 08^h21^m36^s\rightarrow 15^h41^m09^s& 
125.4\rightarrow 235.3& 99/03/21& 51258& 137& 
1.8 - 1.5\cr
& & & & & & & \cr
756& N& 08^h21^m36^s \rightarrow 16^h34^m27^s& 
125.4 \rightarrow 248.6& 99/03/22& 51259& 154& 
1.5 - 1.2\cr
& & & & & & & \cr
}}$$
\end{table}

\begin{table}
\scriptsize
\centerline{Table 2.  SDSS Photometry of Late-Type Dwarfs}
\bigskip
$$\vbox{
\tabskip 1em plus 2em
\halign to \hsize{\hfil$\rm #$\hfil& &   \hfil$\rm#$\hfil \cr
Star&  Run& {\it u^*}& {\it g^*}& {\it r^*}& {\it i^*}& 
{\it z^*}\cr
& & & & & & \cr
& & & & & & \cr
SDSSp~J033035.13-002534.5& 94& 22.33\pm 0.36& 24.09\pm 0.79& 
 {\bf 22.21\pm 0.25}& {\bf 20.11 \pm 0.06}& {\bf 17.98 \pm 0.04}\cr 
& & & & & & \cr
SDSSp~J041320.38-011424.9& 211& 24.63\pm 1.63& 24.69 \pm 0.87&
 {\bf 22.48 \pm 0.21}& {\bf 19.61 \pm 0.03}& {\bf 17.76 \pm 0.02}\cr
& & & & & & \cr
SDSSp~J053951.99-005902.0& 259& 24.59\pm 1.06& 24.67\pm 0.68&
 {\bf 21.49\pm 0.07}& {\bf 19.04\pm 0.02}& {\bf 16.73\pm 0.01}\cr
& & & & & & \cr
SDSSp~J120358.19+001550.3& 756& 24.08\pm 0.06& 24.42\pm 0.39& 
 {\bf 21.31\pm 0.06}& {\bf 18.88\pm 0.01}& {\bf 16.83\pm 0.01}\cr
& & & & & & \cr
SDSSp~J132629.82-003831.5& 752& 23.67\pm 0.33& 24.38\pm 0.44& 
 23.68\pm 0.43& {\bf 21.69\pm 0.14}& {\bf 19.08 \pm 0.05}\cr
& & & & & & \cr
SDSSp~J144001.82+002145.8& 85& 24.13\pm 0.44& 25.06\pm 0.49& 
  {\bf 22.63\pm 0.22}& {\bf 20.47\pm 0.05}& {\bf 18.56\pm 0.03}\cr
& & & & & & \cr
SDSSp~J151547.22-003059.7& 77& 23.21\pm 0.42& {\bf 22.74\pm 0.17}&
 {\bf 21.01 \pm 0.06}& {\bf 18.03 \pm 0.01}& {\bf 16.29 \pm 0.01}\cr
& & & & & & \cr
SDSSp~J163600.79-003452.6& 752& 24.08\pm 0.30& 24.44 \pm 0.40&
  {\bf 21.30\pm 0.06}& {\bf 18.80 \pm 0.01}& {\bf 16.91 \pm 0.01}\cr
}}$$
\noindent Asinh magnitudes (Lupton et al. 1999a) are
quoted; errors are statistical only. Detections are in boldface.
\end{table}

\begin{table}
Table 3. Spectroscopy of SDSS L Dwarfs

\begin{tabular}{lcccccc}\\ \hline \hline
Object & Date & $\rm f_{\lambda} ~(10^{-17} erg~cm^{-2}~s^{-1}~\AA{}^{-1})$&
Spectral Type \\ \hline
& & & \\
SDSS0330$\,-$0025 & 98/11/14& 12.3& L2 \\
SDSS0413$\,-$0014 & 98/12/12 & 13.2& L0 \\
SDSS0539$\,-$0059 & 99/03/20 & 46.1& L5 \\
SDSS1203$\,+$0015 & 99/05/25 & 30.6& L3 \\
SDSS1326$\,-$0038 & 99/05/14 & 3.2 & L8? \\
SDSS1440$\,+$0021 & 99/03/22 & 9.5 & L1 \\
SDSS1515$\,-$0030 & 99/03/15 & 49.7 & M7 \\
SDSS1636$\,-$0034 & 99/05/14 & 32.7& L0 \\
\hline
\end{tabular}

\noindent The flux density in Column 3 is measured at 9500\AA. 
\end{table}

\begin{table}
\scriptsize
\centerline{Table 4.  2MASS Magnitudes for SDSS L Dwarfs}
\bigskip
$$\vbox{
\tabskip 1em plus 2em
\halign to \hsize{\hfil$\rm #$\hfil& &   \hfil$\rm#$\hfil \cr
Object&   2MASS& J& H& K_s& Offset ('')\cr        
& & & & & \cr
& & & & & \cr
SDSS~0330-0025& 2MASSW~J0330351-002534& 
15.29\pm0.05& 14.42\pm0.04& 13.83\pm0.05&
0.1\cr 
& & & & & \cr
SDSS~0413-0114& 2MASSW~J0413204-011424& 
15.33\pm0.05& 14.66\pm0.05& 14.14\pm0.06&
1.1\cr 
& & & & & \cr
SDSS~0539-0059& 2MASSW~J0539520-005901& 
13.99\pm0.03& 13.07\pm0.03& 12.58\pm0.03&
1.0\cr 
          & 2MASSW~J0539519-005901& 14.02\pm0.03& 13.09\pm0.03& 12.51\pm0.03&
	  \cr 
& & & & & \cr
SDSS~1326-0038& 2MASSW~J1326298-003831& 
16.11\pm0.07& 15.04\pm0.06& 14.23\pm0.07&
0.3\cr 
& & & & & \cr
SDSS~1515-0030& 2MASSW~J1515472-003059& 
14.18\pm0.03& 13.58\pm0.03& 13.14\pm0.04&
0.8\cr 
& & & & & \cr
SDSS~1636-0034& 2MASSW~J1636007-003452& 
14.59\pm0.04& 13.93\pm0.04& 13.41\pm0.04&
1.4\cr 
}}$$
\medskip

\noindent SDSS~0539-0059 is in the overlap region between two scans
and was observed twice.
\end{table}

\begin{table}
\scriptsize
\centerline{Table 5.  SDSS-2MASS Colors for SDSS L Dwarfs}
\bigskip
$$\vbox{
\tabskip 1em plus 2em
\halign to \hsize{\hfil$\rm #$\hfil& &   \hfil$\rm#$\hfil \cr
Object& Sp.~T& {\it r^*-i^*}& {\it i^*-z^*}& {\it z^*}-J& J-H& H-K_s& 
{\it i^*}-K_s& {\it z^*}-K_s\cr     
& & & & & & & & \cr
& & & & & & & & \cr
SDSS~1515-0030& M7& 2.98\pm 0.06& 1.74\pm 0.02& 2.11& 0.60& 0.44& 4.89& 3.15\cr
& & & & & & & & \cr
SDSS~0413-0114& L0& 2.46\pm 0.23& 1.80\pm 0.04& 2.50& 0.67& 0.52& 5.47& 3.62\cr
& & & & & & & & \cr
SDSS~1636-0034& L0& 2.50\pm 0.06& 1.89\pm 0.02& 2.32& 0.66& 0.52& 5.39& 3.50\cr
& & & & & & & & \cr
SDSS~0330-0025& L2& 2.10\pm 0.26& 2.13\pm 0.07& 2.69& 0.87& 0.59& 6.29& 4.16\cr
& & & & & & & & \cr
SDSS~0539-0059& L5& 2.45\pm 0.07& 2.31\pm 0.02& 2.74& 0.92& 0.49& 6.46& 4.15\cr
& & & & & & & & \cr
SDSS~1326-0038& L8?& >1.99& 2.61\pm 0.15& 2.97& 1.07& 0.81& 7.46& 4.85\cr
& & & & & & & & \cr
}}$$
\end{table}

\begin{table}
Table 6. Detection Rate of Objects with L Dwarf Colors

\begin{tabular}{lcccccc}\\ \hline \hline
Run &  Number per $\rm deg^2$ \\
\hline \hline
& \\
77&  1/16 \\
85&  1/18 \\
752 & 1/11 \\
756 &  1/12 \\
Orion&  1/9 \\
\hline
\end{tabular}
\medskip

\noindent ``Orion'' refers to those parts of Runs 211, 259, and 273 between right
ascensions of 52$^\circ$ and 76$^\circ$, avoiding regions of extremely
high extinction. 
\end{table}

\begin{figure}
\centerline{\psfig{file=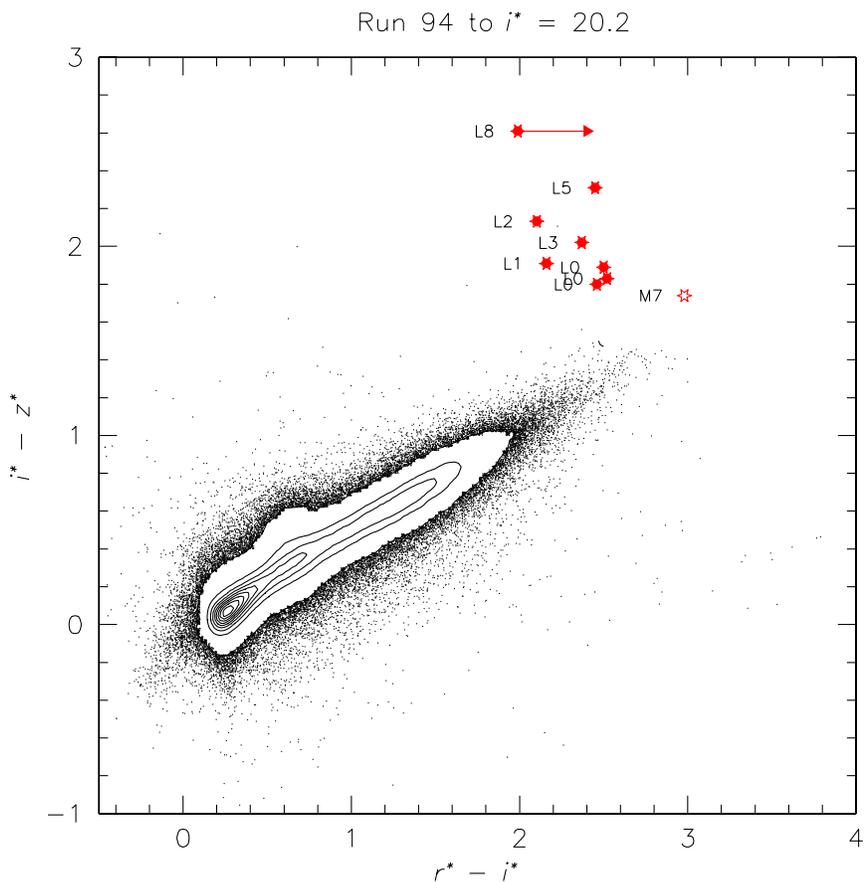,width=12cm}}
\caption{Color-color ($i^*-z^*$ versus $r^*-i^*$) plot for 366859 
point sources brighter than $i^*$ = 20.2 from Run 94, selected to be unblended
(i.e. the image does not overlap the image of another object).  
The contours are drawn every 10\% of the peak density of
points.  The bump at $r^* - i^* \approx 0.7, i^* - z^* \approx 0.4$ is
due to unresolved faint galaxies. 
The colors of the eight objects identified as late-type or substellar
objects in this paper, plus an additional such SDSS object spectroscopically
identified at the Hobby-Eberly Telescope (Schneider \etal\ 1999) are
indicated.  Open symbol: spectral type M.  Filled symbol: spectral type L.
The L8 dwarf is undetected in $r^*$. 
\label{fig1}}
\end{figure}

\begin{figure}
\centerline{\psfig{file=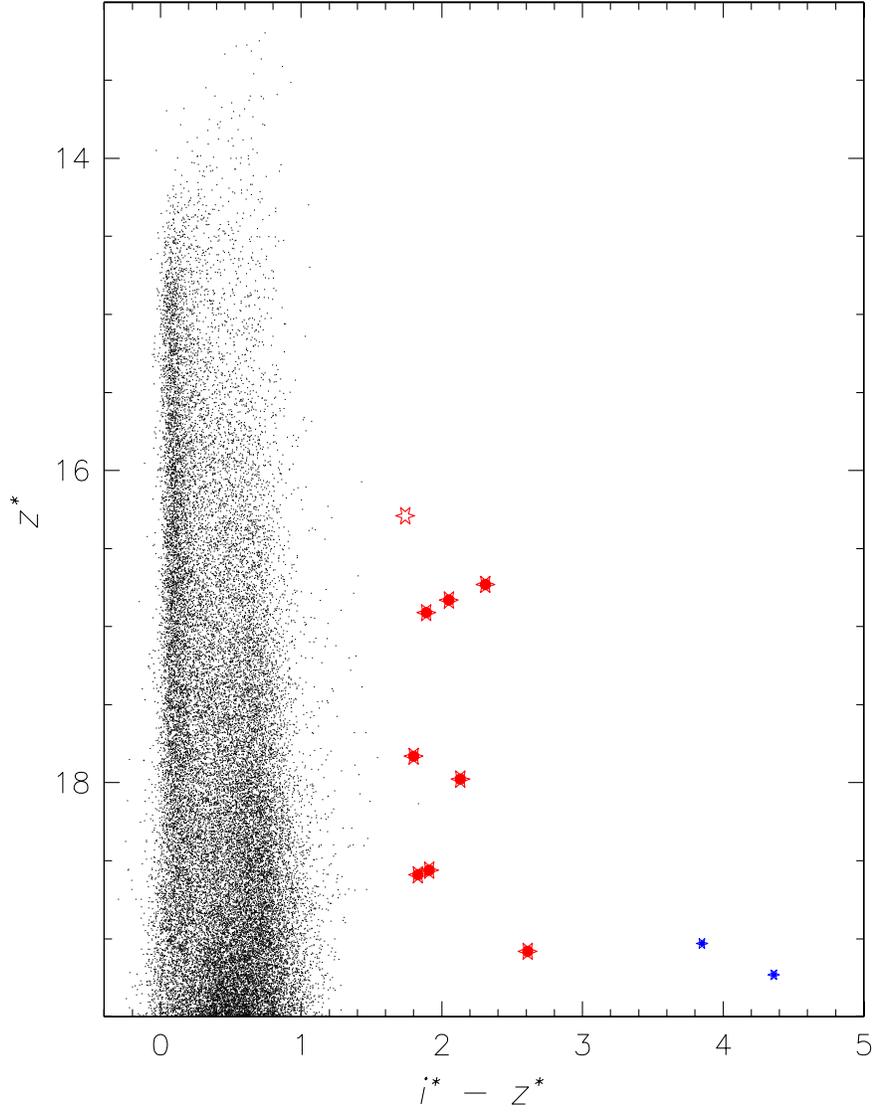,width=12cm}}
\caption{Color-magnitude diagram ($z^*$ vs. $i^* - z^*$)
for 50,000 objects from Run 94 
plus the late type dwarfs
identified by SDSS. 
The M7 dwarf is shown by the large open symbol,
the L dwarfs by large filled symbols, and
the T  dwarfs by small filled symbols.
\label{fig2}}
\end{figure}

\begin{figure}
\centerline{\psfig{file=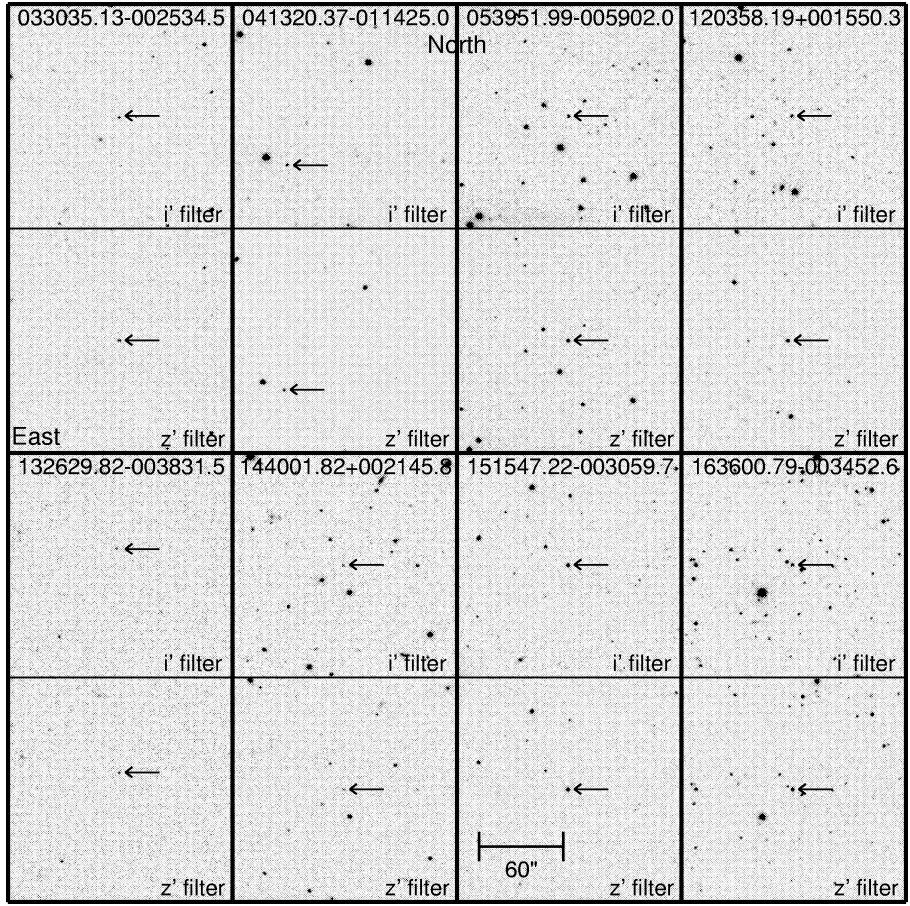,width=12cm}}
\caption{Finding charts in the $i'$ and $z'$ filters for
the M and L dwarfs. The images are $160''$ on a side. \label{fig3}}
\end{figure}

\begin{figure}
\centerline{\psfig{file=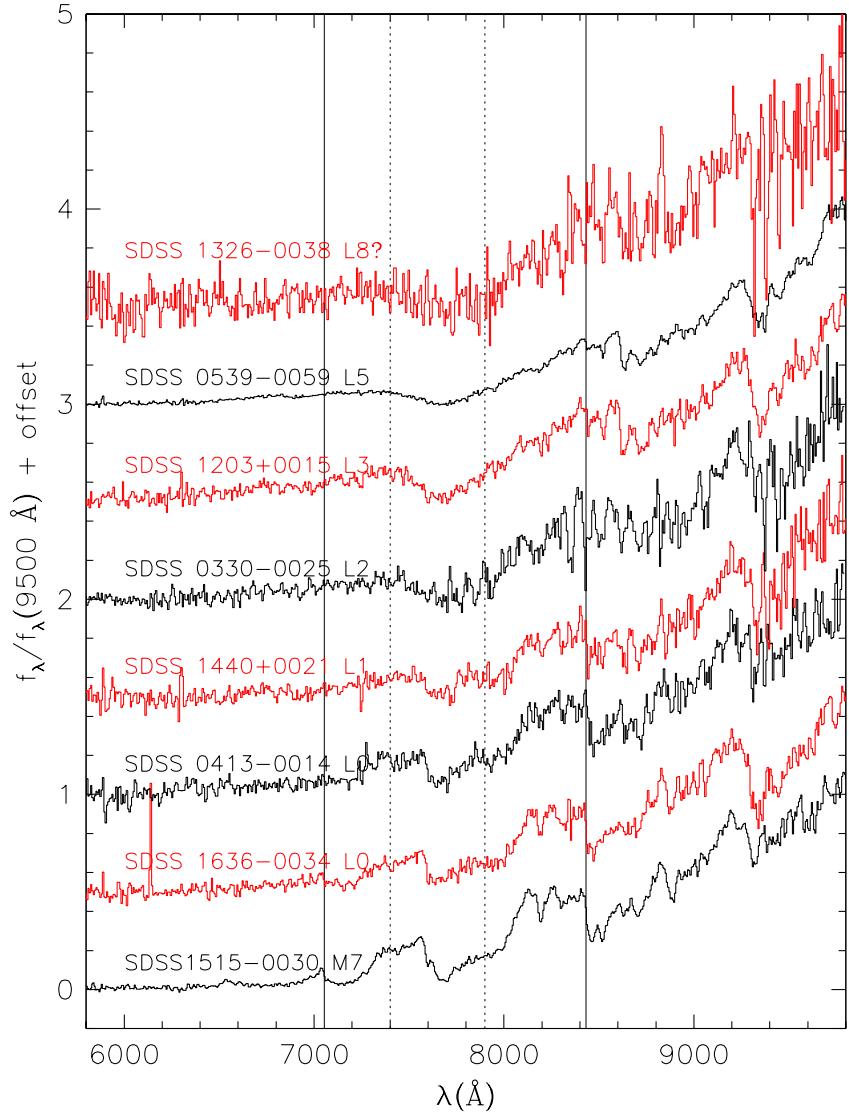,width=12cm}}
\caption{Far-red spectra of the M and L dwarfs, observed by
the Double Imaging Spectrograph (DIS) on the APO 3.5 m telescope.  The
vertical scale for each object has been approximately normalized to the 
flux density of the object at 9500\AA{} (see Table 3), plus an offset.
The objects are
ordered, bottom to top, in increasing spectral type and decreasing effective 
temperature.  The vertical lines show the wavelengths of the TiO 7053 and
8432 \AA{} bandheads (solid lines) and the VO 7400 and 7900 \AA{}
bandheads (dotted lines).
\label{fig4}}
\end{figure}

\begin{figure}
\centerline{\psfig{file=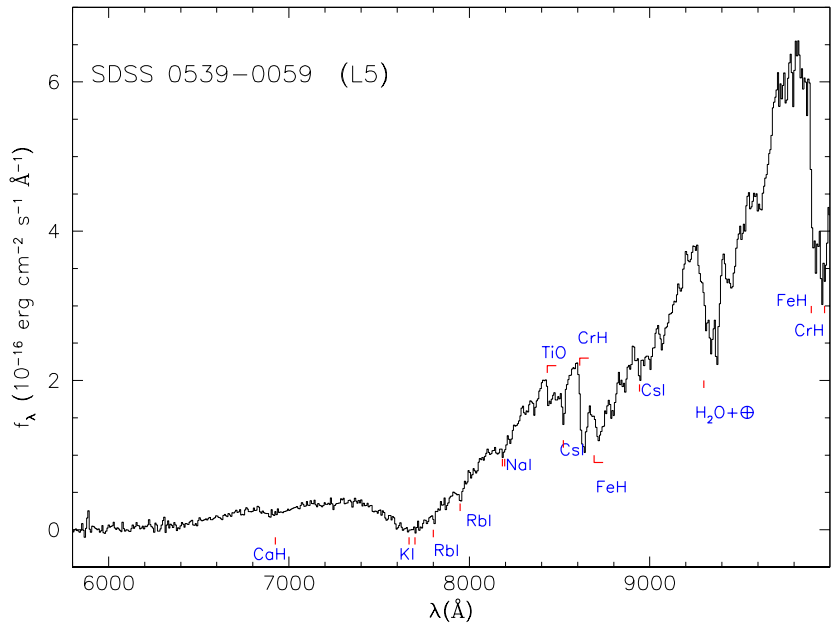,width=12cm,angle=270}}
\caption{Spectra of the brightest L dwarf (L5) found so far by 
SDSS, showing the prominent atomic and molecular absorption features, following
the identifications of Kirkpatrick \etal\ (1999).  Note the large width of
the KI feature.
\label{fig5}}
\end{figure}

\begin{figure}
\centerline{\psfig{file=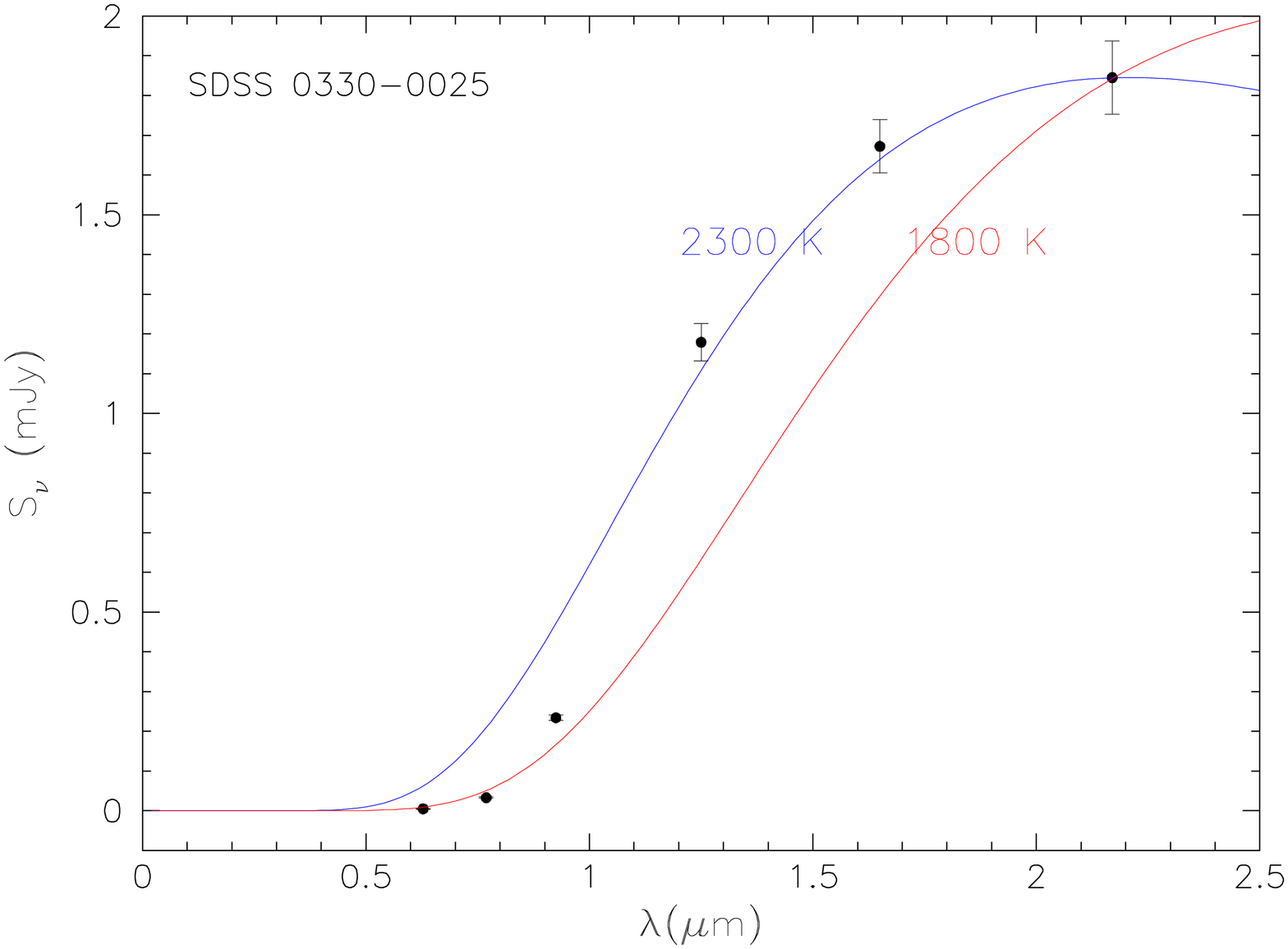,width=12cm,angle=270}}
\caption{Broad-band SED of SDSS 0330-0025 (L2) from SDSS and 2MASS
flux densities.  The two black body curves are normalized to pass through
the 2.2$\rm \mu m$ flux density.
\label{fig6}}
\end{figure}


\begin{references}

\reference{} Burgasser, A.J., Kirkpatrick, J.D., Brown, M.E. \etal\
1999, ApJ, 522, L65


\reference{}Butler, R.P.,
  \& Marcy, G.W. 1998, in ``Brown Dwarfs and Extrasolar
  Planets'', ed.~R. Rebolo, E.L. Mart\'{\i}n \& M.R. Zapatero-Osorio,
  A.S.P. Conference Series 134, 98

\reference{} Cuby, J.G., Saracco, P., Moorwood, A.F.M., D'Odorico, S., 
  Lidman, C., Comer\'on, F., \& Spyromilio, J. 1999, A\&A, in
  press (astro-ph/9907028)

\reference{}Delfosse, X., Tinney, C.G.,
  Forveille, T., Epchtein, N., Bertin, E., Borsenberger, I., Copet, E.,
  de Batz, B., Fouqu\'e, P., Kimeswenger, S., Le Bertre, T., Lascombes, F.,
  Rouan, D., \& Tiphene, D. 1997, A\&A, 327, L25

\reference{}Doi, M. \etal\ 2000, in preparation

\reference{}Fan, X. 1999, AJ, 117, 2528

\reference{}Fan, X., Strauss, M. A., Schneider,
 D.P., \etal\ 1999a, AJ, 118, 1

\reference{}Fan, X., Strauss, M. A., Schneider,
 D.P., \etal\ 1999b, submitted to AJ


\reference{}Fukugita, M., Ichikawa, T., Gunn, J.E.,
Doi, M., Shimasaku, K., \& Schneider, D.P. 1996, AJ, 111, 1748

\reference{} Goldman, B., Delfosse, X., Forveille, T., \etal\ 1999,
  A\&A, in press (astro-ph 9905162)

\reference{} Griffith, C.A., Yelle, R.V, \& Marley, M.S. 1998, Science, 282,
  2063

\reference{}Gunn, J.E., Carr, M.A., Rockosi, C.M., \etal\ 1998, AJ, 116, 3040

\reference{}Gunn, J.E. \& Weinberg D.H. 1995, in
``Wide Field Spectroscopy and the Distant Universe'' ed.~S. Maddox \&
Arag\`on-Salamanca (World Scientific, Singapore), 3

\reference{}Kent, S.M., \etal\ 2000, in preparation

\reference{}Kirkpatrick, J.D., 
  Reid, I.N., Liebert, J., Cutri, R.M., Nelson, B., 
  Beichman, C.A., Dahn, C.C., Monet, D.G., Gizis, J.E., \&
  Skrutskie, M.F. 1999, ApJ, 519, 802 (K99)

\reference{}Liebert, J. 1999, in ``Unsolved Problems
  in Stellar Evolution'', ed.~M. Livio, Cambridge University Press, in
  press (astro-ph/9812061)

\reference{}Lupton, R.H., Gunn,
J.E., \& Szalay, A. 1999, AJ, in press (astro-ph/9903081)

\reference{}Lupton, R.H. {\em et
al.} 2000, in preparation

\reference{}Marcy, G.W.,
  \& Butler, R.P. 1998, in ``Brown Dwarfs and Extrasolar
  Planets'', ed.~R. Rebolo, E.L. Mart\'{\i}n \& M.R. Zapatero-Osorio,
  A.S.P. Conference Series 134, 128

\reference{} Mart\'{\i}n, E.L., Basri, G., Delfosse, X., \&
  Forveille, T. 1997, A\&A, 327, L29

\reference{}Mayor, M., Queloz, D., \& Udry, S.
  1998, in ``Brown Dwarfs and Extrasolar Planets'', ed.~R. Rebolo, E.
  Mart\'{\i}n, \& M.R. Zapatero-Osorio, A.S.P. Conference Series 134, 140

\reference{}Nakajima, T., Oppenheimer, 
  B.R., Kulkarni, S.R., Golimowski, D.A., Matthews, K., \& Durrance,
  S.T. 1995, Nature, 378, 463

\reference{}Oke, J.B., \& Gunn, J.E. 1983, ApJ, 266, 713

\reference{}Oppenheimer, B.R.,
  Kulkarni, S.R., \& Stauffer, J.R. 1999, in `Protostars and Planets
  IV', ed.~V. Mannings, A. Boss \& S. Russell (Tucson: University of 
  Arizona Press) (in press) (astro-ph/9812091)

\reference{}Petravick, D., \etal\ 1994, SPIE, 2198, 935

\reference{}Petravick, D., {\em
et al.} 2000, in preparation

\reference{}
Pier, J.R., Leggett, S.K., Geballe, T.R., et al. 
  \etal\ 1999, in ``Giant Planets and Brown Dwarfs'',
  ed.~M. Marley, A.S.P. Conf. Ser., in press

\reference{}Pier, J. R. \etal\ 2000,
in preparation

\reference{} Queloz, D., Mayor, M., Sivan, J.P., Kohler, D., Perrier,
  C., Mariotti, J.M., \& Beuzit, J.L. 1998, 
  in ``Brown Dwarfs and Extrasolar
  Planets'', ed.~R. Rebolo, E.L. Mart\'{\i}n \& M.R. Zapatero-Osorio,
  A.S.P. Conference Series 134, 324


\reference{}Rebolo, R.,
  Mart\'{\i}n, E.L., \& Zapatero-Osorio, M.R. (eds) 1998, ``Brown Dwarfs
  and Extrasolar Planets'', A.S.P. Conference Series 134.

\reference{}Reid, I.N. \etal\ 1999, ApJ, 521, 613

\reference{}Ruiz, M.T., Leggett, S.K., \& Allard,
  F. 1997, ApJ, 491, L107

\reference{}
Schneider, D.P., Hill, G.J., Fan, X., et al. 1999, submitted to PASP 

\reference{}Siegmund, W. {\em et
al.} 2000, in preparation

\reference{}
Skrutskie, M.F., \etal\ 1997, in ``The Impact of Large-Scale Near
  Infrared Sky Surveys'', eds. F. Garzon et al., Dordrecht:
  Kluwer Academic Publishing Co., 25

\reference{} Smith, J.A., \etal\ 1998, AAS, 193, 02.06

\reference{}
Strauss, M.A., Fan, X., Gunn, J.E., \etal\ 1999, ApJ, 522, 61

\reference{}Tody, D. 1993, in ``Astronomical Data Analysis
and Software Systems II'', ASP Conference Series 52, ed.~R.J. Hanisch,
R.J.V. Brissenden \& J. Barnes, 173

\reference{}
Tsuji, T., Ohnaka, K., \& Aoki, W. 1996, A\&A, 305, L1

\reference{}
Tsuji, T., Ohnaka, K., \& Aoki, W. 1999, ApJ, 520, L119

\reference{}
Tsvetanov, Z.I., Golimowski, D., Zheng, W.  \etal\ 1999, in preparation

\reference{}Tucker, D.L. \etal\ 1998, AAS, 193, 02.08

\reference{}Tucker, D.L. \etal\ 2000, in preparation

\reference{}York, D. \etal\ 2000,
in preparation

\end{references}
\end{document}